\documentclass[12pt]{article}
\usepackage{epsfig,amsfonts,amssymb}
\usepackage{cite,epsf}

\newcommand{\be}{\begin{equation}}
\newcommand{\ee}{\end{equation}}
\newcommand{\ben}{\begin{eqnarray}\displaystyle}
\newcommand{\een}{\end{eqnarray}}

\newcommand{\bea}[1]{\begin{eqnarray}\label{#1} }
\newcommand{\eea}{\end{eqnarray}}

\newcommand{\refb}[1]{(\ref{#1})}

\newcommand{\sectiono}[1]{\section{#1}\setcounter{equation}{0}}

\def\boxempty{{\,\lower0.9pt\vbox{\hrule \hbox{\vrule height 0.25 cm
\hskip 0.25 cm \vrule height 0.25 cm}\hrule}\,}}

\def\one{{\hbox{ 1\kern-.8mm l}}}
\def\zero{{\hbox{ 0\kern-1.5mm 0}}}

\begin{document}
%%%%%%%%%%%%%%%%%%%%%%%%%%%%%%%%%%%%%%%%%%
\begin{titlepage}

\title{
{\Large\bf Fisher Equation for a Decaying Brane}}

\author{
{\large\bf Debashis Ghoshal}\\
{\large\it School of Physical Sciences}\\
{\large\it Jawaharlal Nehru University}\\
{\large\it New Delhi 110067, India}\\
{\tt dghoshal@mail.jnu.ac.in}\\
{}\\
}

\bigskip\bigskip

\date{%August 2008
\begin{quote}
\centerline{\bf Abstract:} {\small We consider the inhomogeneous decay of an 
unstable D-brane. The dynamical equation that describes this process (in 
light-cone time) is a variant of the non-linear reaction-diffusion equation that 
first made its appearance in the pioneering work of (Luther and) Fisher and 
appears in a variety of natural phenomena.}
\end{quote}
}

\bigskip\bigskip
%\leftline{{\bf Report No: HRI-P-08-10-001}} }

%\begin{abstract}
%\end{abstract}

\end{titlepage}
%%%%%%%%%%%%%%%%%%%%%%%%%%%%%%%%%%%%%%%%%%
\maketitle\vfill \eject

%\baselineskip=18pt

%\tableofcontents
%\newpage

%%%%%%%%%%%%%%%%%%%%%%%%%%%%%%%%%%%%%%%%%%
\sectiono{Introduction}\label{introd}
D-branes play an important role in the dynamics of and duality relations between 
different string theories. These are defined by boundary conditions on open strings,
but at the same time are typically stringy non-perturbative excitations of closed 
strings. While some D-branes carry conserved charges and are stable, others are 
unstable. To be precise, all D-branes of different dimensions of the bosonic string 
theory and more than half of those of the type I and II superstring theories are unstable. 
Moreover, configurations of more than one kind of D-branes could be unstable 
even when the constituents are stable individually; for example, a brane-anti-brane
pair. A detailed understanding of this instability provides a valuable window into 
the behaviour of strings. 

Even though a string has an infinite number of vibrational modes, the leading 
effects due to the instability and its characteristic features are governed by the 
lowest mode, the so called tachyonic scalar field. The study of unstable branes 
was pioneered by Sen\cite{Sen:2004nf}, who proposed a set of precise conjectures 
concerning the tachyon. Since then, these conjectures have been checked in the
open string field theory in the level truncation scheme, in various toy models, in 
the boundary string field theory, and finally, an exact solution describing the state
at the (local) minimum of the tachyon potential has been constructed in open 
string field theory. Much less studied, however, is the time-dependent dynamical 
process of tachyon condensation\cite{Moeller:2002vx,Hellerman:2008wp,%
Joukovskaya:2008zv,Barnaby:2008tc,Barnaby:2008pt,Beaujean:2009rb,%
Song:2010hc,Barnaby:2010kx}. 

Ignoring all the details and and technicalities, this is a system which has two
extrema: an unstable maximum (the perturbative vacuum of {\em open} strings) 
and a stable (or locally stable, as in the case of bosonic string) minimum. It is 
expected that the system will make a transition from the unstable to the stable 
phase dynamically. This situation is ubiquitous not only in physics but in various 
other fields, among them biological and chemical systems.  A typical equation that 
governs the dynamics in such cases is the Fisher equation:
\begin{equation}
\partial_t u(t,x) = D\partial_x^2 u(t,x) + r u(t,x)\left(1 - u(t,x)\right),\label{fisher}
\end{equation}
where, $D$ and $r$ are constants; (a more general function $f(u)$ may be 
considered in the RHS). It admits a time-dependent solution that 
corresponds to a {\em front} which separates the two phases (an unstable one 
at $u=0$ and a stable phase at $u=1$) and moves with a characteristic speed 
while retaining its profile. This reaction-diffusion equation and its travelling 
front solution has a long history: Although it was first written by Luther (1906) 
for a chemical system, unaware of this work, Fisher (1937) proposed this 
equation and studied its front solution to describe the spread of an advantageous 
mutation. A detailed mathematical analysis by Kolmogorov, Petrovsky and 
Piskunov was the first in the vast literature\cite{MurrayMB,LDNath} that
followed. The equation of the tachyon field on an unstable D-brane turns out to be 
a variant of this with new elements in the form of time delay and spatial non-locality.   

%%%%%%%%%%%%%%%%%%%%%%%%%%%%%%%%%%%%%%%%%%
\sectiono{Open string field theory}
Henceforth, for definiteness, we shall restrict to the unstable branes of the bosonic string 
theory. The tachyon equation can be obtained from the cubic open  string field 
theory. The string field, expanded in terms of the infinite number of oscillatory modes, is 
\[
\left|\Psi\right> = \left(\phi(X)\,c_1 + \cdots\right) \left|0\right> = \left(\displaystyle{\int}
{d^nk\over (2\pi)^n}\phi(k)\,e^{ik.X}\, c_1 + \cdots\right) \left|0\right>,
\]
where $\phi$ denotes the scalar field corresponding to the lowest mode of the string
and the dots denote the higher excitatations that have been omitted. The cubic action 
of the string field is of the Chern-Simons type:
\[
S_{\mbox{\tiny\rm SFT}} = -{1\over g^2}\left({1\over2}\left<\Psi, Q_B\Psi\right> 
+ {1\over3}\left<\Psi, \Psi\star\Psi \right>\right).
\]
In the above, the products are defined in the conformal field theory on the upper 
half-plane in the usual fashion. If we retain only the tachyon field (level 
truncation to zeroth order) we obtain the action\cite{Moeller:2002vx}:
\begin{equation}
S = -{1\over g^2} \int d^nx \left[{\alpha'\over2}\partial^\mu\phi\partial_\mu\phi - 
{1\over 2}\phi^2 + {K^3\over 3}\, \left(K^{\alpha'\boxempty}
\phi\right)^3\right], \label{phaction}
\end{equation}
where, $K = 3\sqrt{3}/4$. The equation of motion for the tachyon
\begin{equation}
\alpha'\boxempty \phi(t,\mbox{\bf x}) = -\phi(t,\mbox{\bf x}) + 
K^3 e^{\alpha\boxempty}
\left[e^{\alpha\boxempty} \phi(t,\mbox{\bf x})\right]^2,\label{pheom}
\end{equation}
(where, $\alpha = \alpha'\ln K$) contains an infinite number of higher 
derivatives in the interaction term and is, therefore, non-local. 

When $\phi$ depends only on time (spatially homogeneous decay):
\begin{equation}
{d^2\phi(t)\over dt^2} = \phi(t) - K^{3}\, e^{-\alpha\frac{d^2}{dt^2}}
\left[e^{-\alpha \frac{d^2}{dt^2}}\phi(t)\right]^2.
\label{homeqn}
\end{equation}
This equation has solutions\footnote{Empirically, the initial value problem turns 
out to be well-defined with just two initial conditions, say, the position and 
the velocity\cite{Moeller:2002vx}. For further analysis of initial conditions of 
equations of this type, see Refs.\cite{Barnaby:2008tc,Barnaby:2010kx}.} 
that start at the maximum of the potential (at $\phi_U=0$) towards the (local) 
minimum (at $\phi= K^{-3} \simeq 0.456$), but it overshoots and exhibits 
(non-linear) oscillations around the minimum. At late times, these behave wildly.
However, there is not a solution that interpolates between the unstable and the 
stable extrema\cite{Moeller:2002vx}. On the other hand, according to the 
conjectures of Sen, one expects the unstable D-brane to decay into some 
configuration of closed strings, which will carry the energy (density) of the brane.
At the tree level OSFT, however, open strings do not interact with the closed 
string modes for this to happen. 

%%%%%%%%%%%%%%%%%%%%%%%%%%%%%%%%%%%%%
\sectiono{Branes in Linear Dilaton Background}
In order to circumvent this, yet not deal with the complexities of an open-closed 
string field theory, one may consider open string field theory in the presence of 
a closed string background. Perhaps the simplest of these is a linear dilaton 
background\footnote{Another possibility that has cosmological implications is 
to couple the tachyon to gravity\cite{Barnaby:2006hi}.} considered in 
Ref.\cite{Hellerman:2008wp}.  These authors use light-cone coordinates 
$x^\pm=(t \pm x)/\sqrt{2}$, and consider the dilation profile ${\cal D}(x) 
= -D^+x^- \equiv -b x^-$ to study the homogeneous decay of the tachyon as a 
function of light-cone time $x^+$, which we shall call $\tau$ to simplify notation. 
The dilaton, being linear along a null direction, changes the (world-sheet) 
conformal dimension of the tachyon vertex operator $e^{ik.X}$ from $k^2$ to 
$k^2 + ibk^-$ (but does not alter the matter contribution to the central charge). 
Consequently, the equation of motion for the tachyon gets modified from 
Eq.\refb{pheom} to
\begin{equation}
\alpha' \left(b\,{\partial\over\partial\tau} %\phi(x^+,\mbox{\bf x}_\perp) 
- \nabla^2_\perp\right)\phi(\tau,\mbox{\bf x}_\perp) = 
\phi(\tau,\mbox{\bf x}_\perp) - K^3 e^{-2\alpha b\partial_\tau + 
\alpha\nabla^2_\perp} \left[e^{\alpha\nabla^2_\perp} 
\phi(\tau,\mbox{\bf x}_\perp)\right]^2,
\label{pheom-dil}
\end{equation}
where $\mbox{\bf x}_\perp$ denotes the coordinates transverse to the light-cone
coordinates. This is the `Fisher equation for the tachyon on a decaying brane'. 
While there are many variants of the Fisher equation\cite{MurrayMB,LDNath},  
this particular incarnation with a nonlocal nonlinear term and a delayed dependence 
on time, is, to our knowledge, novel.\footnote{Similar forms of nonlocality in 
interactions in biological systems\cite{BFish1,BFish2,Pattern} was pointed out to
me by V.M.~Kenkre.}

The case where $\phi=\phi(\tau)$ depends only on time (homogeneous decay),
and studied in Refs.\cite{Hellerman:2008wp,Barnaby:2008tc,Barnaby:2008pt,%
Beaujean:2009rb,Song:2010hc,Barnaby:2010kx},   
\begin{equation}
\alpha' b\,\partial_+ \phi(\tau) = 
\phi(\tau) - K^3 \left[ \phi(\tau - 2\alpha b)\right]^2,
\label{hutchison}
\end{equation}
is a canonical example of a delayed growth model used, {\em e.g.}, in population
dynamics\cite{MurrayMB}. While the standard logistic growth model has a simple 
interpolating solution, the delay leads to oscillations around the stable fixed point at 
$\phi_S$. This follows from linearizing Eq.\refb{hutchison} 
around the stable fixed\footnote{The linearized equation around the unstable fixed 
point $\phi_U=0$, solved by $\exp\left(\tau/b\right)$, shows the system moving
exponentially away from it, as one would expect of a tachyonic scalar field.} point 
in terms of $\phi=\phi_S+\psi$:
\begin{equation}
\alpha' b \partial_\tau\psi(\tau) = \psi(\tau) - \psi(\tau - 2\alpha b). 
\label{stab_lin}
\end{equation} 
An ansatz on the form $\psi\sim e^{-\lambda\tau}$ leads to a transcendental
equation
\begin{equation}
\alpha' b \lambda + 1 = 2 e^{-2\alpha b\lambda},
\label{transcend}
\end{equation}
that does not have a real solution, but an infinite number of complex solutions
(occurring in complex conjugate pairs). This property is characteristic of delayed 
differential equations. The (oscillatory) convergence to $\phi_S$ is determined 
by the solution with the smallest value of $\mathrm{Re }\,\lambda$. 

%%%%%%%%%%%%%%%%%%%%%%%%%%%%%%%%%%%%%%
\sectiono{The Travelling Front}
Getting back to the Fisher equation \refb{pheom-dil}, let us consider only one
transverse coordinate (denoted by $y$) for simplicity. We seek a travelling 
front solution that moves from the right to the left (so that at any instant of
time the region to the right of the front is converging to the stable fixed
point). First consider the equation linearized around $\phi_U=0$. A 
trial solution of the form $\phi\sim\exp\left(k(y-v(k)\tau\right)$,
leads to the dispersion relation
\begin{equation}
v(k) = {1\over b}\left(k + {1\over k}\right),
\label{lin_disp}
\end{equation}
that has a minimum at $v_{\mathrm{min}} = 2/b$. The wavenumber $k$ is real 
for $v(k)\ge v_{\mathrm{min}}$, therefore, any of these would solve
the linearized equation. For the standard Fisher equation, with a large class of 
nonlinear interactions, the travelling front is proven to select $v_{\mathrm{min}}$
among this\cite{MurrayMB,LDNath}. This feature is likely to be true 
of Eq.\refb{pheom-dil} (with the additional elements of delay and nonlocality)
because the `initial condition'---the leading edge of the wave---is determined by 
the `mass' of the tachyon. However, we shall not dwell on a more rigorous proof
here.

Rather, we look for a travelling front solution in the form of  $\phi(\tau,y) = 
\Phi(\eta=y + v\tau)$, and use singular perturbation 
analysis\cite{MurrayMB,LDNath,SPuri} to determine the solution $\Phi(\eta)$. 
Note, from Eq.\refb{lin_disp}, 
that $v^2b^2\ge 4$. In the absence of a naturally 
small parameter, the idea is to scale $\eta = \sqrt{\varepsilon}\,\xi$ by 
$\varepsilon\equiv 1/v^2b^2 \le 0.25$,  so that the equation takes the form:
\begin{equation}
\partial_\xi\Phi - \frac{\varepsilon}{\sqrt{\alpha'}}\partial_\xi^2\Phi  = \sqrt{\alpha'}
\Phi - \frac{1}{\sqrt{\alpha'}}K^3\,\exp\left(-2\alpha\partial_\xi + \varepsilon\alpha
\partial_\xi^2\right)\left[e^{\varepsilon\alpha\partial_\xi^2}\Phi\right]^2,
\label{RDeq_s_pert}
\end{equation}
Now expand $\Phi(\xi,\varepsilon) = \displaystyle\sum_{n=0}^\infty\Phi_n(\xi) 
\varepsilon^n$ as a power series in $\varepsilon$ and compare terms.
The lowest order equation\footnote{We shall set $\alpha'=1$ from now on.} 
that determines $\Phi_0(\xi)$
\begin{equation}
\partial_\xi\Phi_0 - \Phi_0 + K^3 e^{-2\alpha\partial_\xi}\left[\Phi_0(\xi)\right]^2 = 0,
\label{phi_not_eqn}
\end{equation}
is identical to the homegeneous equation \refb{hutchison}. Therefore, the 
solution of Ref.\cite{Hellerman:2008wp} is a seed for the travelling front. The
correction at ${\cal O}(\varepsilon)$, $\Phi_1(\xi)$ can be solved from
\begin{eqnarray}
{} &{}& 
\partial_\xi\Phi_1 - \Phi_1 +  2 K^3 e^{-2\alpha\partial_\xi}\left[\Phi_0\Phi_1\right]\; = 
\label{phi_one_eqn}\\
{}&{}&
\qquad\qquad\partial_\xi^2\Phi_0 - K^3 e^{-2\alpha\partial_\xi}\left[4\alpha
\Phi_0\partial_\xi^2\Phi_0 + 2\alpha \left(\partial_\xi\Phi_0\right)^2\right], \nonumber
\end{eqnarray}
after substituting $\Phi_0(\xi)$ from Eq.\refb{phi_not_eqn}. The equation at 
${\cal O}(\varepsilon^2)$ is
\begin{eqnarray}
{}&{}&
\partial_\xi\Phi_2 - \Phi_2 + 2 K^3 
e^{-2\alpha\partial_\xi}\left[\Phi_0\Phi_2\right]\;\; = \label{phi_two_eqn}\\
{}&{}&
\qquad\partial_\xi^2\Phi_1 - K^3 
e^{-2\alpha\partial_\xi}\left[4\alpha^2\Phi_0\partial_\xi^4\Phi_0
+ 8\alpha^2 \partial_\xi\Phi_0\partial_\xi^3\Phi_0 + 6\alpha^2\left(\partial^2_\xi
\Phi_0\right)^2\right.\nonumber\\
{}&{}&
\qquad\qquad\qquad\qquad\qquad
\left. 4\alpha\partial_\xi^2\Phi_0\Phi_1 + 
4\alpha\partial_\xi\Phi_0\partial_\xi\Phi_1 +
4\alpha\Phi_0\partial_\xi^2\Phi_1 + \Phi^2_1\right],\nonumber
\end{eqnarray}
by substituting for $\Phi_0(\xi)$ and $\Phi_1(\xi)$ from Eqs.\refb{phi_not_eqn}
and \refb{phi_one_eqn}, respectively, and so on recursively.

\begin{figure}[ht]
\begin{center}
\includegraphics[scale=0.8]{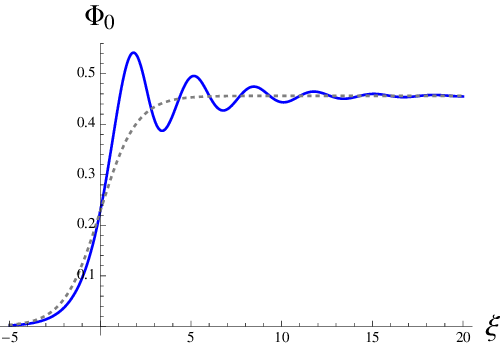}
\hspace*{24pt}
\includegraphics[scale=0.8]{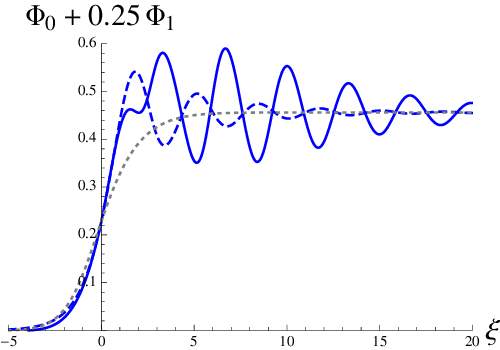}
\end{center}
\caption{{\small
On the left: The leading order solutions $\Phi_0(\xi)$ of the ordinary 
Fisher equation (gray dotted) and the tachyon equation \refb{phi_not_eqn} 
(blue). 
On the right: Solutions upto ${\cal O}(\varepsilon)$---the ${\cal O}(1)$
solution of the tachyon equation is shown as blue dashed line. 
The undulation of the tachyon around the stable vacuum is characteristic
of the delay.}} 
\label{fig:TrFront}
\end{figure}

The functions $\Phi_n$ for the usual Fisher equation can be found analytically. 
For the tachyon, the propagating front can be obtained by {\tt NDSolve} for
delayed differential equation in {\tt mathematica}. We choose an `initial' 
configuration $A e^\xi$ and adjust $A$ such that at $\xi=0$, $\Phi(0)$
is half-way to the stable vacuum at $\phi_S$. Due to the delay, the tachyon
settles to the stable vacuum after damped oscillations. For the correction at
${\cal O}(\varepsilon)$, to tame any unnatural behaviour of the numerical 
solution, we choose an `initial configuration' for $\Phi_1$ that is identical to 
the standard Fisher equation for $\xi<0$. Indeed, in this region, the profile 
of the front in the two cases are rather similar. The results are displayed in 
Fig.\ref{fig:TrFront}.

%%%%%%
%%%%%%%

\begin{figure}[ht]
\begin{center}
\includegraphics[scale=0.8]{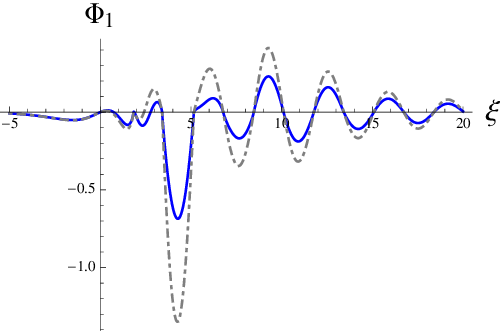}
\hspace*{24pt}
\includegraphics[scale=0.8]{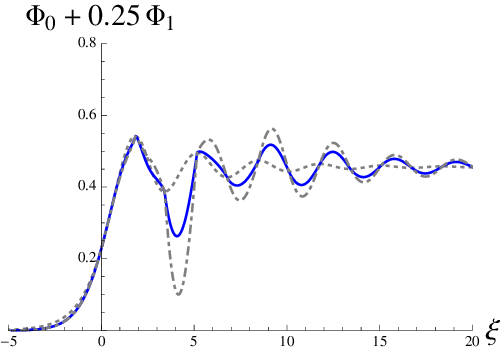}
\end{center}
\caption{{\small
On the left: Solutions at ${\cal O}(\varepsilon)$ obtained by integration of 
Eqs.\refb{phi_one_eqn} (grey dot-dashed) and \refb{alt_phi_one_eqn} (blue). 
On the right: Solutions upto  ${\cal O}(\varepsilon)$. The ${\cal O}(1)$ 
solution is shown as gray dotted line. }} 
\label{fig:TrFrontQuad}
\end{figure}

In adapting singular perturbation theory to our problem, we have naively truncated 
the infinite number of derivatives in $e^{\alpha\partial_\xi^2}$ to a small finite number. 
One could proceed in another way to avoid this problem. We notice that the interaction 
term involves:
\begin{equation}
e^{a\partial_\xi^2} f(\xi) \equiv \mathfrak{G}[f] = \frac{1}{2\sqrt{a\pi}}
\int_{-\infty}^\infty e^{-\frac{(\zeta -\xi)^2}{4a}} f(\zeta) d\zeta,
\label{gausskernel}
\end{equation}
and folding by the Gaussian kernel softens the oscillations. Since, in the limit, $a\to 0$,
the kernel becomes the Dirac $\delta$-function, 
$\mathfrak{dG}[\Phi_0]  = \mathfrak{G}[\Phi_0] - \Phi_0$ is ${\cal O}(\varepsilon)$ in 
our case. The lowest order equation \refb{phi_not_eqn} remains unchanged, but 
the first order correction is now determined by:
\begin{equation}
\partial_\xi\Phi_1 - \Phi_1 + 2 K^3 e^{-2\alpha\partial_\xi}
\left[\Phi_0\Phi_1\right] = g_1(\Phi_0), 
\label{alt_phi_one_eqn}
\end{equation}
where, $g_1(\Phi_0)=\partial_\xi^2\Phi_0 - K^3 e^{-2\alpha\partial_\xi}
\left(2\Phi_0\mathfrak{dG}[\Phi_0] + \mathfrak{dG}[\Phi_0^2]\right)$. Let us find $\Phi_1$ in a
different way: We begin with the standard Fisher equation, which corresponds to 
$\alpha = 0$. In that case, the analogue of the LHS of Eq.\refb{alt_phi_one_eqn} 
(and indeed all the equations at higher orders in $\varepsilon$) are simplified by the 
observation that  $1 - 2 K^3 \Phi_0 = \Phi^{''}_0/\Phi'_0$. This helps to reduce the 
problem of finding $\Phi_{n>0}$ to one of quadrature. Formally, this is still true as 
an operator equation, for a differentiation of Eq.\refb{phi_not_eqn} yields 
$\left[1 - 2 K^3 \left(e^{-2\alpha\partial_\xi}\Phi_0\right) 
e^{-2\alpha\partial_\xi}\right]\partial_\xi\Phi_0 = \partial_\xi^2\Phi_0$. One
can now integrate in {\tt mathematica} to find $\Phi_1$. The constant of
integration is chosen so that the correction vanishes at $\xi=0$\cite{MurrayMB}. 
The solution upto ${\cal O}(\varepsilon)$ obtained this way from 
Eqs.\refb{phi_one_eqn} and \refb{alt_phi_one_eqn} are shown in 
Fig.\ref{fig:TrFrontQuad}. Even though the error seems greater, either due to the 
operator identity not being too accuarte, or due to  numerical integration, the singular 
perturbation method using the integral transform \refb{alt_phi_one_eqn} (instead of 
truncation) clearly shows improved behaviour.

%%%%%%
%%%%%%%

%%%%%%%%%%%%%%%%%%%%%%%%%%%%%%%%%%%%%%
\sectiono{Conclusions}
We close with a couple of comments. The decay of an unstable brane will be 
triggered by the tachyon moving away from the maximum of the potential in a 
finite region of space. In the one dimensional case we have studied, this will 
lead to two fronts, one moving to the left and the other to the right. In higher 
dimensions, in the spherically symmetric case, the Laplacian in 
Eq.\refb{pheom-dil} in the radial variable $r$ does not give a Fisher type 
equation, but will asymptote to one for large $r$\cite{MurrayMB}. 

In summary, the dynamical equation of a tachyon on an unstable D-brane is
a Fisher type reaction-diffusion equation, in which the interaction is smeared
by a Gaussian kernel and is also delayed. It will be interesting to see if these
additional features are useful elsewhere; for example,  its effect on pattern 
formation in biological systems\cite{Pattern} may be worth studying. As for the 
decaying brane, extension of the travelling front to a solution in string field 
theory as well as its stability are among the open problems. 
%%%%%%%%%%%%%%%%%%%%%%%%%%%%%%%%%%%%%%%%
%\bigskip

\noindent{\bf Acknowledgments:} It is a pleasure to thank  Satya Majumdar (who
introduced me to the Fisher equation), Ram Ramaswamy (for discussion and 
Ref.\cite{MurrayMB}) and Anjan Ananda Sen (for help with Mathematica). I have
talked about this subject at various places---thanks are due to the hosts for 
hospitality and to all those in the audience for interesting questions and 
comments. 
%%%%%%%%%%%%%%%%%%%%%%%%%%%%%%%%%%%%

%%%%%%%%%%%%%%%%%%%%%%%%%%%%%%%%%%%%%
\end{document}